# A New Quantum Key Distribution Scheme based on Frequency and Time Coding[*]

ZHU Chang-Hua,　PEI Chang-Xing,　QUAN Dong-Xiao, GAO Jing-Liang　CHEN Nan, YI Yun-Hui

(State Key Laboratory of Integrated Services Networks, Xidian University, Xi'an 710071, China)

*A new scheme of quantum key distribution (QKD) using frequency and time coding is proposed, in which the security is based on the frequency-time uncertainty relation. In this scheme, the binary information sequence is encoded randomly on either the central frequency or the time delay of the optical pulse at the sender. The central frequency of the single photon pulse is set as $\omega_1$ for bit "0" and set as $\omega_2$ for bit "1" when frequency coding is selected. While, the single photon pulse is not delayed for bit "0" and is delayed in $\tau$ for "1" when time coding is selected. At the receiver, either the frequency or the time delay of the pulse is measured randomly, and the final key is obtained after basis comparison, data reconciliation and privacy amplification. With the proposed method, the effect of the noise in the fiber channel and environment on QKD system can be reduced effectively.*

PACS：03.67.Dd, 42.50. Dv

In 1984, a quantum key distribution (QKD) protocol based on Hisenberg uncertainty principle and the no-cloning theorem is proposed by Charles H. Bennett and Gilles Brassard[1]. This protocol, known as BB84, can really provide unconditional security. Since then, the theory and experiments of QKD have got rapid development[2]. Various QKD schemes based on polarization coding[1][3]-[5], phase coding[6]-[10], frequency coding[11]-[13], time coding[14]-[16] and entanglement[2][17][18] were proposed. In these schemes, the one based on polarization coding is influenced by polarization mode dispersion and polarization dependent loss, and the polarization state should be recovered at the receiver for QKD system in optical fiber. In the one based on phase coding the quantum bit error rate (QBER) is related to the interference visibility which is influenced by the noise in the fiber and environment, and the methods to keep the stability of the interference should be adopted. So, the bidirectional auto-compensation structure[8] and differential phase coding schemes[19]-[21] were introduced. The one based on frequency coding needs modulator, e.g. phase modulator or acousto-optic modulator, accordingly is limited by the performance of these devices. In the one based on time coding the coherence of light pulse needs to be tested in order to determine the security of QKD.

[*] Project supported by the National Natural Science Foundation of China(Grant No. 60572147，No.60672119) , the 111 Project (B08038) and the State Key Lab. of Integrated Services Networks (ISN 02080002, ISN 090307).

In this letter, a QKD scheme using frequency coding and time coding is proposed in order to reduce the influence of the channel noise. A similar QKD method based on the energy-time uncertainty relation is also presented[22]. Compared with the scheme in this letter, it is applied to continuous variable system, needs single photon detector with time resolution for time measurement, and also needs many detectors for frequency measurement. So it is high in cost.

The principle of the proposed protocol is shown in Fig.1. At the sender (Alice), there are three lasers, attenuators(Att), coupler, time delay module (TDM), control and driver module(CDM), and personal computer(PC). Laser 1 and laser 2 produce wide pulses (in time domain) with central frequencies $\omega_1$ and $\omega_2$, respectively. Laser 3 produces narrow pulses (in time domain) with central frequency $\omega_3$. Let the frequency spectrum envelope of the lasers 1, 2 and 3 is Gaussian, with the bandwidth as $\sigma_{\omega 1}$, $\sigma_{\omega 2}$ and $\sigma_{\omega 3}$, respectively. Here, the frequency spectrum of the narrow pulse produced by laser 3 covers the frequency spectrum of the pulses produced by laser 1 and laser 2. The wide pulse produced by laser 1 and 2 covers the time duration of the light pulse produced by laser 3 and the delayed one (with time $\tau$). So, the light pulses in the two coding basis are overlayed in time domain and frequency domain. Of course, the pulse width of the lasers is less than coherence length. The time delay of a pulse is induced by TDM. At the receiver (Bob), there are optical switch (OS), circulator(Cir), Fiber Bragg Grating(FBG) filter, Single Photon Detecor(SPD), data collection module(DCM), and personal computer(PC). The frequency components with central frequency $\omega_2$ and bandwidth $\sigma_{\omega 2}$ can be reflected by FBG, while other frequency components of light pulses can propagate through FBG. We name the proposed protocol as FT protocol for short.

The working process of the FT protocol is as follows:

(1)Alice generates two random binary bit sequences: information bit sequence $\{a_n, n=1,2,3,...\}$ and coding bit sequence $\{b_n, n=1,2,3,...\}$. ① If $b_n=0$, then frequency coding is selected. Laser 1 is fired when $a_n=0$, the other two lasers are disabled; laser 2 is fired when $a_n=1$, the other two lasers are disabled. The mean photon number per light pulse is 0.1 after being attenuated. ②if $b_n=1$, then time coding is selected. Laser 3 is fired, the others are disabled. The delay value is 0 when $a_n=0$; the delay value is $\tau$ when $a_n=1$.

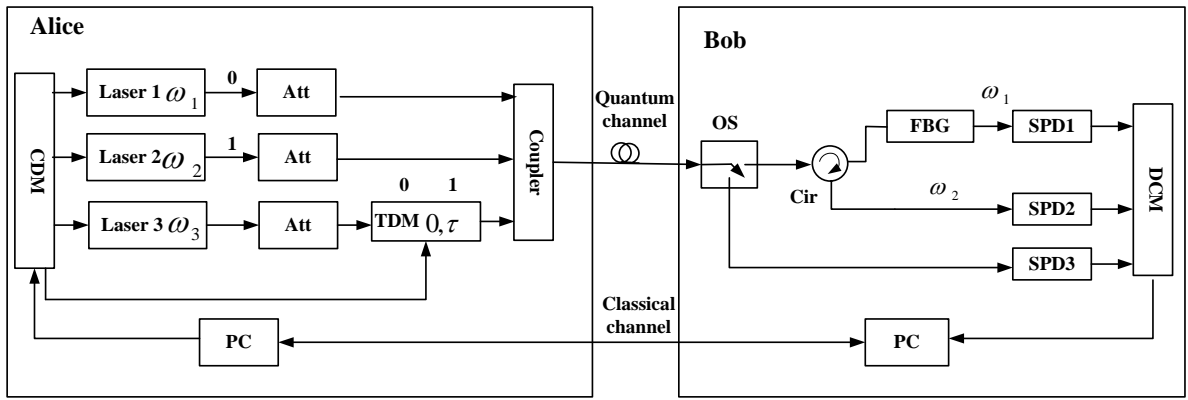

Fig.1 The principle of QKD based on frequency and time coding

(2)The coded light pulse propagates to Bob through optical fiber channel. Bob performs frequency or delay measurement randomly by controlling the optical switch, as shown in Fig.1. ①If frequency measurement is selected, then the photon with central frequency $\omega_2$ and bandwidth $\sigma_{\omega 2}$ is reflected by FBG and is detected by SPD2 with the help of the circular, and the photon with central frequency $\omega_1$ is detected by SPD1. If SPD1 clicks, then the received bit $g_n=0$; if SPD2 clicks, then the received bit $g_n=1$. ②If time measurement is selected, the photon is detected by SPD3. The received bit $g_n=0$ when delay is 0, the received bit $g_n = 1$ when delay is $\tau$. ③Let $g_n = -1$ whether all the three detectors don't click or photon count occurs under the two basis at the same time, that is, SPD1 or SPD2 clicks and SPD3 also clicks.

(3) Bob tells Alice the measurement types (frequency or time delay) by public calssical channel.

(4) Alice tells Bob the measurement types which are consistant with the coding types and keeps the element of $\{a_n, n = 1, 2, 3, ...\}$ with same types as the sifted key sequence $\{s_m, m = 1, 2, 3, ...\}$.

(5) Bob keeps the element of the received sequence $\{g_n, n = 1, 2, 3, ...\}$ with the same types of measurement and coding as the sifted key $\{r_m, m = 1, 2, 3, ...\}$.

(6) Alice compares part bits of $\{s_m, m = 1, 2, 3, ...\}$ and part bits of $\{r_m, m = 1, 2, 3, ...\}$ with Bob by calssical channel. Then she calcaulate QBER. If $QBER > QBER_{th}$, then eavesdropper (Eve) exits, QKD falses and return to (1). Otherwise, performs (7).

(7) Alice and Bob obtain the final key after data reconcilliation and privacy amplification by classical channel.

The security of the FT protocol can be analyzed elementarily as follows: (1) since the weak coherent light pusles coded in frequency and time delay overlay each other in frequency domain and time domain, the states in the two bases are non-orthogonal and can not be discriminated correctly. In fact, Eve can not measure the frequency and time delay of the light pulse accurately and simutaneuosly because of frequency-time uncertainty relation, $\sigma_\omega \sigma_t \geq 1$, where $\sigma_\omega$ and $\sigma_t$ is the width of frequency spectrum and time[23]. (2) For intercept-resend attack, Eve can obtain the information bit correctly with a probability of 50% when the measurement type of Eve is different with the coding type, e.g. when Alice performs frequency coding, Eve measures the time delay and he can obtain 0 or 1 with the same probability for the light pulse with central frequencies $\omega_1$ or $\omega_2$ and time interval covering $0 \sim \tau$; when Alice performs time coding, Eve measures the frequency and he can obtain correct result with a probability of 50% for the light pulse with frequency bandwidth covering $\omega_1$ and $\omega_2$. So, Bob can detect the Eve's attack easily by QBER estimation. (3) Trojan attack can not be implemented for the light pulse of QKD propagates along the fiber in one-way.

The feasibility of the FT protocol is mainly determined by the devices that meet with the parameters requirement of the system. Let the pulse width of the lasers 1, 2 and 3 in time domain is $\sigma_{t1}$, $\sigma_{t2}$ and $\sigma_{t3}$, respectively. We set $\omega_1$ and $\omega_2$ as the telecommunication band (1550nm), $\omega_3 = (\omega_1 + \omega_2)/2$. The parameters have the following relations: $\sigma_{\omega 3} \geq \sigma_{\omega 1} + \sigma_{\omega 2}$, $\sigma_{t1} \geq 2\sigma_{t3}$, $\sigma_{t1} = \sigma_{t2}$, $\sigma_{\omega 1} = \sigma_{\omega 2}$. The delay $\tau$ should be $\sigma_{t1} - \sigma_{t3} \geq \tau \geq \sigma_{t3}$ for correct discrimination of the pulse and the delayed one. In this system, $\tau$ should be larger than trigging interval of single photon detector. Recently, single photon detector based on superconductor has the trigging frequency over GHz and is low in dark count[24]. The lasers that meet with these requirements can be obtained in the market.

Now, we analyze the quantum bit error rate (QBER) and final key rate of the FT protocol. Let $\mu$ denote the mean photon number of each weak coherent pulse. The photon number $n$ in each pulse is poisson distributed, $p_n = \frac{\mu^n}{n!} e^{-\mu}$. Let $\alpha$ ($dB/Km$) denote the fiber loss, $l$ ($Km$) denote the fiber length. Then the transmission rate of the fiber $t_f = 10^{-\alpha l/10}$. If the transmission rate of the quantum channel at the receiver is $t_B$, the efficiency of the detector is $\eta_D$, then the whole transmission rate of the system $t = t_f t_B \eta_D$. Assume that the transmission of the photons in each pulse is independent. So, the count rate at the receiver for the pulse with n photons is $t_{n,nd} = 1-(1-t)^n$ when dark count is not taken into account, and the count rate is $t_n = t_{n,nd} + mp_d(1-t_{n,nd})$ when dark count is taken into account, where $p_d$ is the dark count probability, $m$ is the number of the detectors in each basis. From the statistical point of view, when dark count is not taken into account the mean count rate $R_{nd}$ of weak coherent pulse at the receiver is $R_{nd} = \sum_{n=0}^{\infty} p_n t_{n,nd} = 1 - e^{-\mu t}$. When dark count is taken into account the mean count rate $R$ of weak coherent pulse is $R = \sum_{n=0}^{\infty} p_n t_n = 1 - e^{-\mu t} + mp_d e^{-\mu t}$. So, the sifted key rate $R_{sift}$ can be given as

$$R_{sift} = R \cdot p_s = \left(1 - e^{-\mu t} + mp_d e^{-\mu t}\right) p_s \tag{1}$$

Where, $p_s$ denotes the percent of qubit used to sift key.

Here, QBER means the error ratio of the information bit which the receiver obtains after basis comparison, denoted as $Q$. Qubit error comes from two ways, one is the incorrect clicks of the detectors because of channel noise, the other is the dark count of detectors. Based on these QBER can be given as

$$Q = \frac{\left[R_{nd} p_{opt} + \frac{m}{2}(1-R_{nd})p_d\right]p_s}{R_{sift}} \qquad (2)$$

Where, $p_{opt}$ denotes the probability with which photons arrive at incorrect detectors, $\frac{1}{2}$ is the error probability from dark count, $p_{s\_FT} = \frac{1}{2}$, $m=2$. So the sifted key rate $R_{sift\_FT}$ is $R_{sift\_FT} = \left[1-e^{-\mu t} + 2p_d e^{-\mu t}\right] p_{s\_FT}$. For the FT protocol, let $\sigma_{t1}(\sigma_{t2})$ be $1 \sim 1.2 ns$, $\sigma_{t3}$ be $500 \sim 600 ps$, the associated bandwidth $\sigma_{\omega1}(\sigma_{\omega2}) = 1.047 \sim 1.257 \times 10^{10} rad/s$, $\sigma_{\omega3} = 2.094 \sim 2.513 \times 10^{10} rad/s$, and the detection gate duration be twice of the width of the pulse sent. So, the effect of time spread and frequency spread from dispersion on detection results can be omitted, that is $p_{opt} \approx 0$ when the basis is the same, QBER $Q_{FT}$ of the QKD system can be given as

$$Q_{FT} = \frac{(1-R_{nd})p_d p_s}{R_{sift\_FT}} \qquad (3)$$

For BB84 QKD system, in which polarization coding or phase coding based on M-Z interferometer is applied, $p_{s\_BB84} = \frac{1}{2}$. The sifted key rate of BB84 QKD system $R_{sift\_BB84} = \left[1-e^{-\mu t} + 2p_d e^{-\mu t}\right] p_{s\_BB84}$, the QBER can be given as[13]

$$Q_{BB84} = \left[R_{nd}\frac{1-V}{2} + (1-R_{nd})p_d\right]\frac{p_{s\_BB84}}{R_{sift\_BB84}} \qquad (4)$$

In equation (4), $V$ denotes the visibility, $0 \leq V \leq 1$. For phase coding, $V$ denotes the visibility of the interference meter. For polarization coding, $V$ denotes the degree to which polarization rotation because of polarization mode dispersion affects the measurement results.

In the proposed protocol, the weak coherent light pusles coded in frequency and time delay are in two non-orthogonal states, the error and information obtained by Eve is removed after data reconciliation and privacy amplification, being the same as the BB84 protocol, and the final secure key rate can be given as[2]

$$R_{net} = R_{sift} \left( I_{AB} - I_{AE} \right) \tag{5}$$

Where, $I_{AB}$ denotes the mutual information between Alice and Bob, and $I_{AB} = 1 - H_2(Q)$. $H_2(Q)$ is the binary entropy, $H_2(Q) = -Q\log_2(Q) - (1-Q)\log_2(1-Q)$. $I_{AE}$ denotes the mutual information between Alice and Eve. We assume that Eve perform photon number splitting (PNS) attack first and no error occurs, that is, the photons are split to Eve with a proportion $1 - t_f$ and the residual photons are transmitted to Bob with a proportion $t_f$ through a channel without loss. Hence, the mean photon number of the light pulse obtained by Eve is $\mu(1-t_f)$ and Eve can get the information with the probability $\mu(1-t_f)$ from each pulse sent by Alice. Then, Eve performs intercept-resend attack against the pulses in which the detectors of Eve don't click in PNS attack.

Since the visibility is the ratio of the subtraction of the number of photon count by the incorrect detector from the number of photon count by correct detector over the total number of photon count, the incurred bit error rate to Bob can be given as $\frac{1-V}{2}$ under the assumption of the photon counting of the incorrect detector derives from intercept-resend attack. For BB84 and FT protocol, if the probability with which Eve performs intercept-resend attack is $p_{IR}$, the error rate is $0.25 p_{IR}$ and the information Eve obtains is $0.5 p_{IR}$ [25], so the information Eve obtains can be given as $1-V$. The mutual information between Alice and Eve $I_{AE\_FT} = \mu(1-t_f) + (1-V)$, the final key rate can be given as follows

$$R_{net\_FT} = R_{sift\_FT} \left[ 1 - H_2(Q_{FT}) - I_{AE\_FT} \right] \tag{6}$$

Accordingly, for BB84 protocol with polarization coding or phase coding, the mutual information between Alice and Eve is $I_{AE\_BB84} = \mu(1-t_f) + (1-V)$ [14]. Then the final key rate is

$$R_{net\_BB84} = R_{sift\_BB84} \left[ 1 - H_2(Q_{BB84}) - I_{AE\_BB84} \right] \tag{7}$$

Let $t_B = 1$, $\eta_D = 0.1$, $\alpha = 0.25$, $p_d = 10^{-5}$, V is 1, 0.9, 0.8 and 0.7 respectively. The relations of the final key rates to the distance are shown in Fig.2. As shown in Fig.2, when $V = 0.7$, the communication distance of FT protocol is increased by 65Km. When $V = 0.8$, the distance is increased by 12Km. When $V < 0.7$, the scheme with polarization coding and phase coding can not work, while the scheme with the FT protocol can work correctly. So the scheme with frequency and time coding can work in worse visibility compared with BB84 protocol using polarization coding or phase coding.

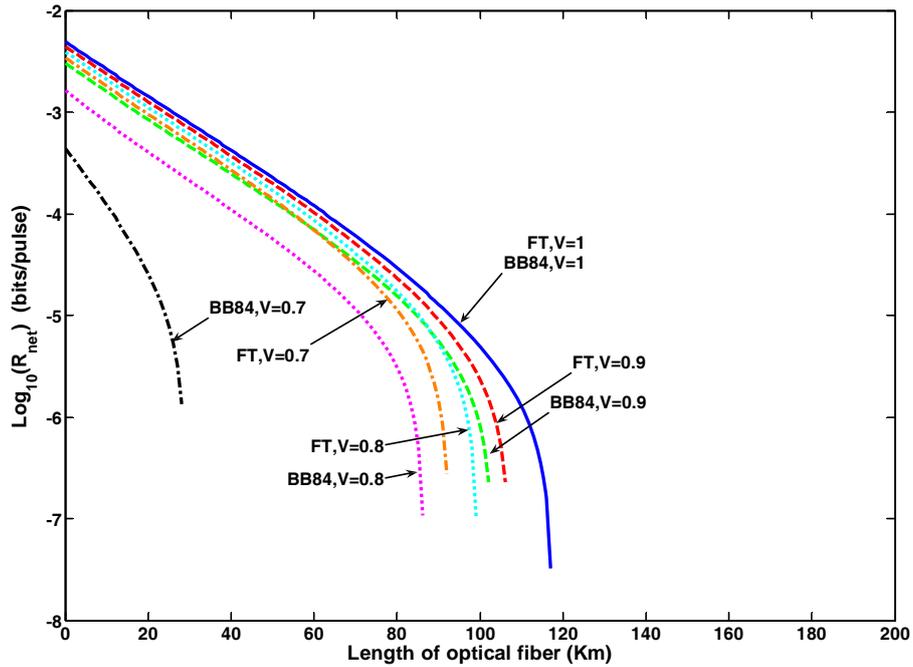

Fig.2 The relation of final key rate and length of fiber

In conclusion, based on time-frequency uncertainty principle a new QKD scheme using frequency and time coding is proposed. Compared with polarization coding and phase coding, the effect of birefringence and interference instability on the performance of QKD system is reduced. The analysis results of final key rates show that the FT protocol can work at worse visibility. A rigorous proof of the security of the FT protocol remains a work in progress. Furthermore, the theory of decoy states can be applied to increase the distance further and overcome the shortcomings by weak coherent pulse [26][27].